\newcommand{\reskg}     {\mbox{$0.88^{+0.09}_{-0.08}$}}

\newcommand{\resgz}     {\mbox{$0.987^{+0.034}_{-0.033}$}}
\newcommand{\reslam}    {\mbox{$-0.060^{+0.034}_{-0.033}$}}
\newcommand{\resgfz}    {\mbox{$-0.04^{+0.13}_{-0.12}$}}
\newcommand{\ddkgOA}    {\mbox{$ 0.029$}}
\newcommand{\ddkgIS}    {\mbox{$ 0.006$}}
\newcommand{\ddkgFR}    {\mbox{$ 0.038$}}
\newcommand{\ddkgTG}    {\mbox{$ 0.011$}}
\newcommand{\ddkgDE}    {\mbox{$ 0.006$}}
\newcommand{\ddkgBG}    {\mbox{$ 0.015$}}
\newcommand{\ddkgEB}    {\mbox{$ 0.008$}}
\newcommand{\ddkgBE}    {\mbox{$ 0.001$}}
\newcommand{\ddkgCR}    {\mbox{$ 0.004$}}
\newcommand{\ddkgTO}    {\mbox{$ 0.053$}}
\newcommand{\shadkg}    {\mbox{$-0.12^{+0.09}_{-0.08}$}}
\newcommand{\comdkg}    {\mbox{$-0.12^{+0.09}_{-0.08}$}}
\newcommand{\dkgint} {\mbox{$[-0.27,\;\; 0.07]$}}
\newcommand{\ddgzOA}    {\mbox{$ 0.011$}}
\newcommand{\ddgzIS}    {\mbox{$ 0.002$}}
\newcommand{\ddgzFR}    {\mbox{$ 0.013$}}
\newcommand{\ddgzTG}    {\mbox{$ 0.005$}}
\newcommand{\ddgzDE}    {\mbox{$ 0.003$}}
\newcommand{\ddgzBG}    {\mbox{$ 0.004$}}
\newcommand{\ddgzEB}    {\mbox{$ 0.004$}}
\newcommand{\ddgzBE}    {\mbox{$ 0.000$}}
\newcommand{\ddgzCR}    {\mbox{$ 0.003$}}
\newcommand{\ddgzTO}    {\mbox{$ 0.019$}}
\newcommand{\shadgz}    {\mbox{$-0.013^{+0.034}_{-0.033}$}}
\newcommand{\comdgz}    {\mbox{$-0.013^{+0.034}_{-0.033}$}}
\newcommand{\dgzint}    {\mbox{$[-0.077,\;\; 0.054]$}}
\newcommand{\dlamOA}    {\mbox{$ 0.010$}}
\newcommand{\dlamIS}    {\mbox{$ 0.003$}}
\newcommand{\dlamFR}    {\mbox{$ 0.018$}}
\newcommand{\dlamTG}    {\mbox{$ 0.004$}}
\newcommand{\dlamDE}    {\mbox{$ 0.005$}}
\newcommand{\dlamBG}    {\mbox{$ 0.011$}}
\newcommand{\dlamEB}    {\mbox{$ 0.005$}}
\newcommand{\dlamBE}    {\mbox{$ 0.004$}}
\newcommand{\dlamCR}    {\mbox{$ 0.005$}}
\newcommand{\dlamTO}    {\mbox{$ 0.026$}}
\newcommand{\shalam}    {\mbox{$-0.061^{+0.035}_{-0.034}$}}
\newcommand{\comlam}    {\mbox{$-0.060^{+0.034}_{-0.033}$}}
\newcommand{\lamint} {\mbox{$[-0.13,\;\; 0.01]$}}
\newcommand{\dgfzOA}    {\mbox{$ 0.031$}}
\newcommand{\dgfzIS}    {\mbox{$ 0.003$}}
\newcommand{\dgfzFR}    {\mbox{$ 0.047$}}
\newcommand{\dgfzTG}    {\mbox{$ 0.005$}}
\newcommand{\dgfzDE}    {\mbox{$ 0.013$}}
\newcommand{\dgfzBG}    {\mbox{$ 0.012$}}
\newcommand{\dgfzEB}    {\mbox{$ 0.017$}}
\newcommand{\dgfzBE}    {\mbox{$ 0.005$}}
\newcommand{\dgfzCR}    {\mbox{$ 0.000$}}
\newcommand{\dgfzTO}    {\mbox{$ 0.061$}}
\newcommand{\shagfz}    {\mbox{$-0.04^{+0.13}_{-0.12}$}}
\newcommand{\comgfz}    {\mbox{$-0.04^{+0.13}_{-0.12}$}}
\newcommand{\gfzint} {\mbox{$[-0.28,\;\; 0.21]$}}
\newcommand{\qqkgst}    {\mbox{$ 0.18^{+0.38}_{-0.20}$}}
\newcommand{\qqdgst}    {\mbox{$ 0.08^{+0.09}_{-0.07}$}}
\newcommand{\qqlast}    {\mbox{$ 0.00^{+0.09}_{-0.07}$}}
\newcommand{\qqgfst}    {\mbox{$-0.41^{+0.20}_{-0.21}$}}

\newcommand{\qlkgst}    {\mbox{$-0.23^{+0.09}_{-0.08}$}}
\newcommand{\qldgst}    {\mbox{$-0.031^{+0.035}_{-0.034}$}}
\newcommand{\qllast}    {\mbox{$-0.075^{+0.036}_{-0.035}$}}
\newcommand{\qlgfst}    {\mbox{$ 0.15^{+0.16}_{-0.16}$}}

\newcommand{\lvkgst}    {\mbox{$ 0.31^{+0.34}_{-0.27}$}}
\newcommand{\lvdgst}    {\mbox{$-0.05^{+0.11}_{-0.11}$}}
\newcommand{\lvlast}    {\mbox{$-0.09^{+0.09}_{-0.08}$}}
\newcommand{\lvgfst}    {\mbox{$ 0.12^{+0.48}_{-0.44}$}}

\newcommand{\qqkgsy}    {\mbox{$ 0.17^{+0.65}_{-0.26}$}}
\newcommand{\qqdgsy}    {\mbox{$ 0.09^{+0.11}_{-0.09}$}}
\newcommand{\qqlasy}    {\mbox{$ 0.01^{+0.28}_{-0.14}$}}
\newcommand{\qqgfsy}    {\mbox{$-0.37^{+0.24}_{-0.25}$}}

\newcommand{\qlkgsy}    {\mbox{$-0.23^{+0.09}_{-0.08}$}}
\newcommand{\qldgsy}    {\mbox{$-0.023^{+0.038}_{-0.037}$}}
\newcommand{\qllasy}    {\mbox{$-0.062^{+0.038}_{-0.037}$}}
\newcommand{\qlgfsy}    {\mbox{$ 0.10^{+0.17}_{-0.16}$}}

\newcommand{\lvkgsy}    {\mbox{$ 0.30^{+0.35}_{-0.28}$}}
\newcommand{\lvdgsy}    {\mbox{$-0.06^{+0.11}_{-0.11}$}}
\newcommand{\lvlasy}    {\mbox{$-0.09^{+0.09}_{-0.08}$}}
\newcommand{\lvgfsy}    {\mbox{$ 0.10^{+0.49}_{-0.45}$}}

\newcommand{\qqkgex}    {\mbox{$\pm 0.20$}}
\newcommand{\qqdgex}    {\mbox{$\pm 0.07$}}
\newcommand{\qqlaex}    {\mbox{$\pm 0.08$}}
\newcommand{\qqgfex}    {\mbox{$\pm 0.20$}}

\newcommand{\qlkgex}    {\mbox{$\pm 0.12$}}
\newcommand{\qldgex}    {\mbox{$\pm 0.032$}}
\newcommand{\qllaex}    {\mbox{$\pm 0.033$}}
\newcommand{\qlgfex}    {\mbox{$\pm 0.15$}}

\newcommand{\lvkgex}    {\mbox{$\pm 0.29$}}
\newcommand{\lvdgex}    {\mbox{$\pm 0.13$}}
\newcommand{\lvlaex}    {\mbox{$\pm 0.09$}}
\newcommand{\lvgfex}    {\mbox{$\pm 0.34$}}

\newcommand{\qqkgcs} {\mbox{$ 11.3/15$}}
\newcommand{\qqdgcs} {\mbox{$  9.2/15$}}
\newcommand{\qqlacs} {\mbox{$ 11.0/15$}}
\newcommand{\qqgfcs} {\mbox{$  8.8/15$}}
\newcommand{\qlkgcs} {\mbox{$ 10.3/15$}}
\newcommand{\qldgcs} {\mbox{$ 13.6/15$}}
\newcommand{\qllacs} {\mbox{$ 13.3/15$}}
\newcommand{\qlgfcs} {\mbox{$ 14.7/15$}}
\newcommand{\lvkgcs} {\mbox{$ 14.6/15$}}
\newcommand{\lvdgcs} {\mbox{$ 12.9/15$}}
\newcommand{\lvlacs} {\mbox{$  9.8/15$}}
\newcommand{\lvgfcs} {\mbox{$ 14.9/15$}}
\newcommand{\cokgcs} {\mbox{$ 43.4/53$}} 
\newcommand{\codgcs} {\mbox{$ 39.4/53$}} 
\newcommand{\colacs} {\mbox{$ 36.2/53$}} 
\newcommand{\cogfcs} {\mbox{$ 42.0/53$}}
 
\newcommand{\smkgcs} {\mbox{$ 45.1/54$}} 
\newcommand{\smdgcs} {\mbox{$ 39.5/54$}} 
\newcommand{\smlacs} {\mbox{$ 39.2/54$}} 
\newcommand{\smgfcs} {\mbox{$ 42.1/54$}}

\newcommand{\LepII}{\mbox{LEP}}
\newcommand{\LepI}{\mbox{LEP}}

\newcommand{\Pythia}{\mbox{PYTHIA}}
\newcommand{\Herwig}{\mbox{HERWIG}}
\newcommand{\Ariadne}{\mbox{ARIADNE}}
\newcommand{\Excalibur}{\mbox{EXCALIBUR}}

\newcommand{\YFSww}{\mbox{YFSWW3}}
\newcommand{\Koralw}{\mbox{KORALW}}
\newcommand{\Koralz}{\mbox{KORALZ}}
\newcommand{\Bhwide}{\mbox{BHWIDE}}

\newcommand{\Ftogen}{\mbox{F2GEN}}

\newcommand{\KK}{\mbox{KK2F}}
\newcommand{\Com}{centre-of-mass}
\newcommand{\Grace}{\mbox{\tt grc4f}}

\newcommand{\SM}{Standard Model}
\newcommand{\MC}{Monte Carlo}
\newcommand{\tgc}{{\small TGC}}

\newcommand{\GeV}{\mbox{$\mathrm{GeV}$}}

\newcommand{\GeVcc}{\mbox{$\mathrm{GeV}\!/\!{\it c}^2$}}
\newcommand{\Ipb}{\mbox{pb$^{-1}$}}
\newcommand{\beq}{\begin{equation}}
\newcommand{\eeq}{\end{equation}}
\newcommand{\bea}{\begin{eqnarray}}
\newcommand{\eea}{\end{eqnarray}}
\newcommand{\ra}{\mbox{$\rightarrow$}}

\def\gappeq{\mathrel{ \rlap{\raise.5ex\hbox{$>$}}
                      {\lower.5ex\hbox{$\sim$}}  } }
\def\lappeq{\mathrel{ \rlap{\raise.5ex\hbox{$<$}}
                      {\lower.5ex\hbox{$\sim$}}  } }
\newcommand{\Mw}{\mbox{$M_{\mathrm{W}}$}}

\newcommand{\twsq}{\mbox{$\tan^2\theta_w$}}

\newcommand{\epem}{\mbox{$\mathrm{e^+e^-}$}}

\newcommand{\lplm}{\mbox{$\ell\overline{\ell}$}}
\newcommand{\Zz}{\mbox{${\mathrm{Z}^0}$}}
\newcommand{\WW}{\mbox{$\mathrm{W^+W^-}$}}
\newcommand{\Wm}{\mbox{$\mathrm{W^-}$}}
\newcommand{\Wp}{\mbox{$\mathrm{W^+}$}}

\newcommand{\qq}{\mbox{$\mathrm{q\overline{q}}$}}
\newcommand{\Qqll}{\qq\lplm}

\newcommand{\qqqq}{\mbox{$\qq\qq$}}
\newcommand{\Wqq}{\mbox{$\mathrm{q\overline{q} }$}}

\newcommand{\lnu}{\mbox{$\ell\overline{\nu}_{\ell}$}}
\newcommand{\lnubar}{\mbox{$\overline{\ell}^\prime\nu_{\ell^\prime}$}}

\newcommand{\enu}{\mbox{$\mathrm{e\overline{\nu}_{e}}$}}
\newcommand{\mnu}{\mbox{$\mu\overline{\nu}_{\mu}$}}
\newcommand{\tnu}{\mbox{$\tau\overline{\nu}_{\tau}$}}

\newcommand{\Wenu}{\mbox{$\epem \rightarrow \mathrm{W}\enu$}}

\newcommand{\Zee}{\mbox{$\epem\rightarrow\Zz\epem$}}
\newcommand{\ZZ}{\mbox{$\epem\rightarrow\Zz\Zz$}}

\newcommand{\ZGqq}{\mbox{$\Zz/\gamma\rightarrow\qq$}}

\newcommand{\roots}{\mbox{$\sqrt{s}$}}

\newcommand{\Zgamma}{\mbox{$\Zz/\gamma$}}

\newcommand{\al}{\mbox{$\alpha$}}
\newcommand {\ee} {\mbox{$\mathrm{e}^+ \mathrm{e}^-$}}

\newcommand {\mm} {\mbox{$\mu^+ \mu^-$}}
\newcommand {\nn} {\mbox{$\nu\overline{\nu}$}}
\newcommand {\tautau} {\mbox{$\tau^+ \tau^-$}}

\newcommand {\eeee}   {\ee\ra\ee}
\newcommand {\eemumu} {\ee\ra\mm}
\newcommand {\eetautau} {\ee\ra\tautau}
\newcommand {\eenunu} {\ee\ra\nn}

\newcommand{\Cthw}{\mbox{$\cos\theta_{\mathrm{W}} $}}

\newcommand{\Cthstl}{\mbox{$\cos \theta_\ell^{*}$}}
\newcommand{\Phistl}{\mbox{$\phi_\ell^{*}$}}
\newcommand{\Cthstj}{\mbox{$\cos\theta_{\scriptscriptstyle \mathrm{jet}}^{*}$}}
\newcommand{\Phistj}{\mbox{$\phi_{\scriptscriptstyle \mathrm{jet}}^{*}$}}

\newcommand{\Cthsto}{\mbox{$\cos \theta_1^{*}$}}
\newcommand{\Phisto}{\mbox{$\phi_1^{*}$}}
\newcommand{\Cthstt}{\mbox{$\cos \theta_2^{*}$}}
\newcommand{\Phistt}{\mbox{$\phi_2^{*}$}}
\newcommand{\OO}{\mbox{${\cal O}$}}

\newcommand{\Lnln}{\lnu\lnubar}
\newcommand{\Qqln}{\Wqq\lnu}
\newcommand{\Qqen}{\Wqq\enu}
\newcommand{\Qqmn}{\Wqq\mnu}
\newcommand{\Qqtn}{\Wqq\tnu}
\newcommand{\Qqqq}{\Wqq\Wqq}

\newcommand{\gz}{\mbox{$g_1^{\mathrm{z}}$}}
\newcommand{\gfz}{\mbox{$g_5^{\mathrm{z}}$}}
\newcommand{\kg}{\mbox{$\kappa_\gamma$}}
\newcommand{\kz}{\mbox{$\kappa_{\mathrm{z}}$}}
\newcommand{\lgg}{\mbox{$\lambda_\gamma$}}
\renewcommand{\lg}{\mbox{$\lambda$}}
\newcommand{\lz}{\mbox{$\lambda_{\mathrm{z}}$}}
\newcommand{\dgz}{\mbox{$\Delta g_1^{\mathrm{z}}$}}

\newcommand{\dkg}{\mbox{$\Delta \kappa_\gamma$}}

\newcommand{\dkz}{\mbox{$\Delta \kappa_{\mathrm{z}}$}}

\newcommand{\LL}{\mbox{$\log L$}}

\newcommand {\xse}{cross-section}

\newcommand{\MOO}{\mbox{$\overline{\cal O}$}}
\newcommand{\OOEXP}[2]{\mbox{$E[{\cal O}_{#1}^{#2}]$}}

\documentstyle[epsfig,a4p,cite,12pt]{article}

\begin{document}
\bibliographystyle{plain}
\begin{titlepage}
\begin{center}{\large   EUROPEAN ORGANISATION FOR NUCLEAR RESEARCH
 }\end{center}\bigskip
\begin{flushright}
CERN-EP/2003-042 \\
July 14, 2003  \\
\end{flushright}
\bigskip\bigskip\bigskip\bigskip\bigskip
\begin{center}
 {\LARGE\bf \boldmath 
Measurement of charged current triple gauge boson couplings using
W pairs at LEP}
\end{center}
\bigskip\bigskip
\begin{center}{\Large The OPAL Collaboration}
\end{center}\bigskip
\bigskip
\bigskip\begin{center}{\large  Abstract}\end{center}
Triple gauge boson couplings are measured from W-pair 
events recorded by the OPAL detector at LEP at \Com\ energies 
of 183 -- 209 GeV with a total integrated luminosity of 680~\Ipb.
Only CP-conserving couplings are considered and SU(2)$\times$U(1) 
relations are used, resulting in four independent couplings, \kg, 
\gz, \lgg\ and \gfz. Determining each coupling in a separate fit,
assuming the other couplings to take their  \SM\ values, we obtain  
\kg=\reskg, \gz=\resgz, \lgg=\reslam\ and \gfz=\resgfz, where the 
errors include both statistical and systematic uncertainties. 
Fits are also performed allowing some of the couplings to vary 
simultaneously. All results are consistent with the \SM\ predictions.
\bigskip\bigskip
\bigskip\bigskip\bigskip
\begin{center}
{\large Submitted to the European Physical Journal C}
\end{center}

\end{titlepage}


\begin{center}{\Large        The OPAL Collaboration
}\end{center}\bigskip
\begin{center}{
G.\thinspace Abbiendi$^{  2}$,
C.\thinspace Ainsley$^{  5}$,
P.F.\thinspace {\AA}kesson$^{  3}$,
G.\thinspace Alexander$^{ 22}$,
J.\thinspace Allison$^{ 16}$,
P.\thinspace Amaral$^{  9}$, 
G.\thinspace Anagnostou$^{  1}$,
K.J.\thinspace Anderson$^{  9}$,
S.\thinspace Arcelli$^{  2}$,
S.\thinspace Asai$^{ 23}$
% ...
}\end{center}\bigskip
\bigskip
$^{  1}$School of Physics and Astronomy, University of Birmingham,
Birmingham B15 2TT, UK
\newline
$^{  2}$Dipartimento di Fisica dell' Universit\`a di Bologna and INFN,
I-40126 Bologna, Italy
\newline
$^{  3}$Physikalisches Institut, Universit\"at Bonn,
D-53115 Bonn, Germany
\newline
% ...
$^{  5}$Cavendish Laboratory, Cambridge CB3 0HE, UK
\newline
% ...
$^{  9}$Enrico Fermi Institute and Department of Physics,
University of Chicago, Chicago IL 60637, USA
\newline
% ...
$^{ 16}$Department of Physics, Schuster Laboratory, The University,
Manchester M13 9PL, UK
\newline
% ...
$^{ 22}$Department of Physics and Astronomy, Tel Aviv University,
Tel Aviv 69978, Israel
\newline
$^{ 23}$International Centre for Elementary Particle Physics and
Department of Physics, University of Tokyo, Tokyo 113-0033, and
Kobe University, Kobe 657-8501, Japan
\newline
% ...
\newpage


\section{Introduction}
 \label{sec:intro}

 The triple gauge boson vertices, WW$\gamma$ and WWZ, are expected 
in the \SM\ due to its non-Abelian nature. The \SM\ predicts exact values
of the associated parameters, the triple gauge boson couplings 
(\tgc s). Any significant deviation of the measured values from this 
prediction would be a signature of new physics beyond the \SM.
Several measurements of these couplings  have already been performed
at LEP using W-pair~\cite{tgc161-analysis,tgc172-analysis,
tgc183-analysis,tgc189-analysis,OTHERLEPWW-tgc}, 
single W~\cite{OTHERSW} and single photon~\cite{OTHERSG} production.
Limits on \tgc s also exist from studies of di-boson production at the
Tevatron~\cite{Tevatron-tgc}. This paper reports results on \tgc s
using W-pair events from the full OPAL LEP2 data sample. 

Assuming a general Lorentz-invariant Lagrangian which satisfies
electromagnetic gauge invariance, charge conjugation (C) and 
parity (P) invariance, five independent \tgc\ parameters are expected
to describe the WW$\gamma$ and WWZ vertices.
These can be taken as \gz, \kz, \kg, \lz\ and 
\lgg~\cite{LEP2YR,HAGIWARA,BILENKY,GAEMERS}.
In the \SM\, \gz=\kz=\kg=1 and \lz=\lgg=0.
When the restriction on C and P invariance is dropped,
but their product, CP, is required to be invariant, a sixth 
coupling, \gfz~\cite{LEP2YR,HAGIWARA,BILENKY,GAEMERS}, which 
violates both C and P, is added. This coupling vanishes in the \SM.
Couplings which violate CP~\cite{wpol-analysis} are not 
considered in this analysis.

Precision measurements on the \Zz\ resonance
and lower energy data motivate the following
SU(2)$\times$U(1) relations between the five C and P conserving 
couplings~\cite{LEP2YR,BILENKY},
\begin{eqnarray*}
\label{su2u1}
\dkz & = & -\dkg \twsq +  \dgz, \\ 
\lz  & = &  \lgg. 
\end{eqnarray*} 
Here $\Delta$ indicates a deviation of the respective quantity 
from its \SM\, value and $\theta_w$ is the weak mixing angle at
tree level, defined by $\cos\theta_w=\Mw/M_{\rm Z}$.
These two relations leave three independent C and P conserving
couplings, $\dkg$, $\dgz$ and \lg(=\lgg=\lz), plus the C and P 
violating coupling \gfz.  These couplings are not significantly 
restricted~\cite{DERUJULA,HISZ} by measurements from data collected
at the \Zz\ pole at \LepI\ and SLC.

While the single W and single photon processes can also be used 
to measure \tgc s, W-pair production is by far the most sensitive 
process for this measurement. Its
sensitivity comes from the triple gauge boson vertex connecting
the intermediate $s$-channel \Zz/$\gamma$ to the outgoing W-bosons.
The \tgc\ contribution depends on the helicity
states of the outgoing W bosons, determining 
the angular distributions of the W bosons and of their 
decay products. The total W-pair
\xse\ is also affected by the \tgc s, but for \Com\ energies well
above the threshold for W-pair production and for small values of 
anomalous couplings its sensitivity, compared with the
angular distributions, is much lower. Very often, however,
the angular distribution analysis yields two ambiguous
solutions for the \tgc s, one being quite far away from the
Standard Model values~\cite{SEKULIN}, and the information from the
total \xse\ can help in resolving this ambiguity.  

The production and decay of W-pair events can be characterized 
by five angular variables. 
These are the \Wm\ production polar angle\footnote{ 
The OPAL right-handed coordinate system is defined 
such that the origin is at the centre of the detector,
the $z$-axis is parallel to, and has positive sense along, the e$^-$ beam 
direction, $\theta$ is the polar angle with respect to $z$ and $\phi$ is 
the azimuthal angle around $z$ with respect to the $x$-axis, which
points to the centre of the LEP ring.}, $\theta_W$, 
the polar and azimuthal angles, $\theta^*$ and $\phi^*$, of the   
decay fermion from the \Wm\ in the \Wm\ rest frame\footnote{
The axes of the right-handed coordinate system in the W 
rest-frame are defined such that the $z$-axis is along
the parent W flight direction in the overall centre-of-mass 
system, and the $y$-axis is in the direction 
$\overrightarrow{e^-} \times \overrightarrow{W}$ where 
$\overrightarrow{e^-}$ is the electron beam 
direction and $\overrightarrow{W}$ is the parent W flight direction.},
and the corresponding polar and azimuthal angles of the decay anti-fermion
from the \Wp. The experimental accessibility of these angles depends
on the final state produced when the W bosons decay.

All W-pair final states are used in this study, namely the 
leptonic, \lnu\lnubar, the semileptonic, \Wqq\lnu, and the 
hadronic, \Wqq\Wqq\ final states,
with branching fractions of 10.5\%, 43.9\% and 45.6\% 
respectively in the \SM.


The next section discusses the OPAL data and 
Monte Carlo samples. The following three sections then 
present the first part of this study using the W-pair angular 
distributions. For this purpose, the kinematic observables of each
event are reconstructed as described in Section~\ref{sec:recon}
and then used in Section~\ref{sec:oo} to construct optimal observables
and analyse them in terms of the \tgc s. The study of systematic errors 
for this part of the analysis is summarised in Section~\ref{sec:sys}.
Section~\ref{sec:rate} describes the second part of this analysis, 
where the total W-pair event rate is used to extract additional 
information on the \tgc s. 
The \tgc\ results are presented in Section \ref{sec:combtgc} 
and summarised in the last section. 

\section{Data and Monte Carlo samples}
\label{sec:data}
The OPAL detector is described in detail elsewhere~\cite{OPAL}.
The data were collected during 1997-2000 around eight different 
\Com\ energies. The integrated luminosity at each energy is evaluated 
using small angle Bhabha scattering events observed in the silicon
tungsten forward calorimeter. 
The luminosity-weighted average values of the \Com\ 
energies and the corresponding luminosities are listed in 
Table~\ref{tab:shape}.   

The main \MC\ generator used to simulate the W-pair signal is KandY, 
which is a combination of 
\Koralw1.51~\cite{KandY} and \YFSww~\cite{YFSWW} running concurrently.
These programs generate all four-fermion final states using the full set of
electroweak diagrams including the \WW\ production diagrams 
(class\footnote{
In this paper, the three lowest order W-pair production diagrams, 
{\em i.e.} $t$-channel $\nu_{\mathrm{e}}$ exchange and $s$-channel 
\Zgamma\ exchange, are referred to as ``CC03'', following the 
notation of \cite{LEP2YR}.} 
CC03) and other four-fermion graphs, such as \Wenu, \Zee\ and \ZZ.
All diagrams are corrected for initial and final state radiation,
while a more complete correction of \OO(\al) radiative effects is
applied to the CC03 diagrams. For each centre-of-mass energy, 500,000 
events have been generated using these \MC\ programs, and 
these events are used as our reference samples for the \tgc\ analysis.
The \tgc\ dependence is obtained by reweighting these samples 
using the four-fermion matrix element calculation taken from
the \Excalibur~\cite{EXCALIBUR} \MC\ program. This 
program has also been used to generate samples with anomalous
\tgc s which are used to cross-check our \tgc\ extraction method.
Although \Excalibur\ includes less complete \OO(\al) radiative effects, 
it can still be used for small anomalous \tgc\ values, where the 
effect of the missing \OO(\al) corrections on the anomalous \tgc\ 
contribution can be neglected.

Background sources to the W-pair signal are four-fermion final states,
such as \Qqll, which are produced only by diagrams that do not involve
the triple gauge boson vertex, two-fermion final states and two-photon
processes. The four-fermion background is generated  
using \Koralw1.42\ \cite{KORALW} and \Grace~\cite{GRC4F}. As an 
alternative for systematic studies we replace \Koralw1.42\ by
\Excalibur. Two-fermion final states are generated using 
\KK~\cite{KK2f}. For systematic studies, \Koralz~\cite{KORALZ}, 
for \eemumu, \eetautau\ and \eenunu, and \Bhwide\cite{BHWIDE}, for 
\eeee, are also used. Background to the semileptonic final state from 
single-tag two-photon processes, where the outgoing electron or positron
is detected, is evaluated using \Herwig~\cite{HERWIG}, and  
\Ftogen~\cite{F2GEN} is used as an alternative for systematic studies. 
Background from leptonic final states from untagged two-photon 
processes, where both outgoing electron and positron escape undetected
into the beam pipe, is calculated using the Vermaseren~\cite{VERMASEREN}
and BDK~\cite{BDK} generators. 

To estimate the fragmentation and hadronisation systematics, \MC\ 
samples of \WW\ events were produced by the \Koralw1.42 generator,
with the fragmentation and hadronisation stages generated separately 
by either \Pythia5.7~\cite{PYTHIA}, \Herwig6.2~\cite{HERWIG} or 
\Ariadne4.08~\cite{ARIADNE}. These hadronic simulation generators
have been tuned to \Zz\ hadronic decays~\cite{TUNE}. Similarly, 
for the \ZGqq\ background, the same hadronic simulation programs have 
been used to fragment and hadronise events produced by the \KK\ 
generator.


All \MC\ samples mentioned above
were processed by the full OPAL simulation program~\cite{GOPAL} 
and then reconstructed in the same way as the data.


\section{W-pair event selection and reconstruction}
\label{sec:recon}
  W-pair events are selected with the same procedure used in the 
W-pair cross-section measurement as described in 
references~\cite{tgc183-analysis,sigww189}. There are three
independent selections corresponding to the three final states.
Each candidate is then reconstructed in order to obtain the maximum 
possible information on the W production and decay angles, which are 
needed to extract the couplings. Events which cannot be well 
reconstructed are rejected from the sample, as will be described below.
The numbers of candidates left at the different \Com\ energies for 
the three final states
are listed, along with the expected values, in Table~\ref{tab:shape}.

  Three kinematic fits with different sets of requirements are used in 
the event reconstruction:
\begin{itemize}
\item[A.] Require conservation of energy and momentum, neglecting 
Initial State Radiation (ISR).
\item[B.] Additionally constrain the reconstructed masses of the two 
W-bosons to be equal.
\item[C.] Additionally constrain each reconstructed W mass to 
the average measured value from the Tevatron\footnote{
The \LepII\ results for the W mass are not used for the \tgc\ measurement, 
since they have been obtained under the assumption that W pairs are 
produced according to the \SM, whereas W production at 
the Tevatron does not involve the triple gauge boson vertex.},
\Mw=80.456 \GeVcc~\cite{MWTEV}. 
\end{itemize}

\begin{table}[thbp]
 \begin{center}
 \begin{tabular}{||c|r||r|r||r|r||r|r||} \hline
\roots & $\int{\cal L}dt$ & \multicolumn{2}{|c||}{\Qqln\ events} & 
 \multicolumn{2}{|c||}{\Qqqq\ events} & 
 \multicolumn{2}{|c||}{\Lnln\ events} \\ \cline{3-8}
(GeV) & (\Ipb) & Observed & Expected & Observed & Expected
 & Observed & Expected \\ \hline 
182.7 & 57.4 &  328  &  331.1 & 408 & 418.9 &  32 &  37.4 \\
188.6 & 183.0 & 1090 & 1123.5 &1437 &1388.3 & 130 & 124.3 \\
191.6 & 29.3 &  168  &  182.8 & 223 & 222.8 &  19 &  19.9 \\
195.5 & 76.4 &  513  &  478.9 & 637 & 594.3 &  55 &  52.5 \\
199.5 & 76.6 &   451 &  479.7 & 557 & 585.0 &  52 &  52.1 \\
201.6 & 37.7 &  230  &  236.3 & 296 & 286.1 &  32 &  25.3 \\
204.9 & 81.6 &  475  &  512.4 & 578 & 606.5 &  47 &  53.2 \\
206.6 & 136.5 & 899  &  854.0 &1051 &1012.3 &  92 &  90.0 \\ \hline
all   & 678.5 & 4154 & 4198.7 &5187 &5114.2 & 459 & 454.7 \\ \hline
 \end{tabular}
\caption{Observed and expected numbers of data candidates
 selected for the angular distribution analysis after all cuts
in the different final states and for different \Com\ energies.} 
\label{tab:shape}
 \end{center}
\end{table}

For \qqqq\ events, where all four final state fermions are measurable, 
fits A, B and C have 4, 5 and 6 constraints respectively.
For \Qqen\ and \Qqmn\ events the number of constraints is reduced by 3 
due to the invisible neutrino. For \Qqtn\ events there 
is at least one additional unobserved neutrino from the $\tau$ decay,
resulting in a loss of one constraint.
The momentum sum of the track(s) assigned to the $\tau$ can still
be used as an approximation of the $\tau$ flight direction, relying
on its high boost, but the $\tau$ energy is left unknown. Finally for
\Lnln\ events, where none of the leptons is a $\tau$, there are two
invisible neutrinos. Hence, six constraints are lost and only 
requirement C is applied.

  In the following we discuss the reconstruction of each final state
separately.


\subsection{Reconstruction of \boldmath{\Qqln} final states}
\label{subs:recqqln}
 
Starting from the sample used for the W-pair \xse\ measurement,
we select candidate \Qqen\ and \Qqmn\ events with a  
reconstructed lepton track. The track charge is needed
to reconstruct the \Wm\ polar angle. 
For the \Qqtn\ events, either one track or a narrow jet 
consisting of three tracks is assigned as the $\tau$ decay 
product.

  The OPAL tracking detectors are used to reconstruct the muon
momentum and the electron direction, whereas the electromagnetic 
calorimeters are used to give a more accurate measurement 
of the electron energy. As explained above, the direction of $\tau$ 
candidates is given directly by the $\tau$ decay products, whilst 
the energy of the $\tau$ is estimated using a kinematic fit.   

The remaining tracks and 
calorimeter clusters in the event are grouped into two jets
using the Durham $k_\perp$ algorithm~\cite{DURHAM}. The total 
energy and momentum of each of the jets are calculated using
the charged track momenta and calorimeter energies and 
correcting for double counting employing the method described 
in~\cite{MT}.   

\begin{table}[thbp]
\begin{center}
\begin{tabular}{|l|c|c|l|c|} \hline
Class & Final & \% of & Requirements on fits A, B, C & 
Kinem. variables \\
      & state & events &          & taken from \\ \hline
a) & \Qqen, \Qqmn &30.4 & $P_C\ge 0.20$ & fit C \\ 
b) & \Qqen, \Qqmn &10.1 & $0.05\le P_C<0.20$ & fit C \\ 
c) & \Qqen, \Qqmn & 6.9 & $0.01\le P_C<0.05$ & fit C \\ 
d) & \Qqen, \Qqmn & 4.7 & $0.001\le P_C<0.01$ & fit A \\ 
e) & \Qqen, \Qqmn &10.0 & $P_C<0.001$ & fit A \\ 
f) & \Qqen, \Qqmn & 8.3 & C failed, $P_A\ge 0.001$ & fit A \\    
g) & \Qqen,\Qqmn & 1.2 & C failed, A failed or $P_A<0.001$ & 
  directly measured \\    
h) & \Qqtn        &21.7 & $P_B\ge0.20$ & fit B \\ 
i) & \Qqtn        & 6.8 & $0.025\le P_B<0.20$ & fit A \\ \hline
\end{tabular}
\caption{Definition of the nine quality classes of \Qqln\ events.
$P_A$, $P_B$ and $P_C$ are the probabilities of kinematic fits A, B 
and C.} 
\label{tab:classes}
\end{center}
\end{table}

The different kinematic fits described above are used to improve
the resolution
in the five kinematic variables used for the \tgc\ analysis.
This resolution is correlated with the kinematic fit probabilities
which indicate the event reconstruction quality. Therefore we use  
the results of the kinematic fits to classify the
events into nine quality classes which are used later in 
the \tgc\ analysis. The exact definitions of these classes and
the fraction of accepted events in each class are
listed in Table~\ref{tab:classes}. The event population in each
class is well modelled by the \MC. To optimize the resolution
in the kinematic variables used in the \tgc\ analysis they are 
taken, according to the quality class, either from the directly 
measured values or from one of the three kinematic fits, as detailed 
in Table~\ref{tab:classes}. The \Qqtn\ events which fail kinematic
fit B or pass it with fit probability below 0.025 are rejected.
This cut suppresses those \Qqtn\ events which are 
correctly identified as belonging to this decay channel 
but where the $\tau$ decay products are not identified correctly, 
leading to an incorrect estimate of the $\tau$ flight direction or 
its charge. The fraction of such events in the \Qqtn\ sample
is thus reduced from 18\% to 12\%.

Finally, the kinematic variables are
used to calculate the optimal observables (see Section~\ref{sec:oo})
which are used to extract the \tgc s. Those events (typically 1\% 
of the sample) with optimal observable
values in the far tails of their distributions are interpreted
as being badly measured and discarded. 

Performing the complete analysis on \MC\ samples generated with
different anomalous couplings shows that our event selection does 
not introduce any bias on the \tgc\ results. 

The purities of the final samples are about 96\%, 98\% and 91\% 
for the \Qqen, \Qqmn\ and \Qqtn\ final states respectively. 
The background consists mainly of \tgc-independent four-fermion 
final states and \ZGqq\ events. 
Cross migration between the \Qqtn\ and each of the \Qqen\ and 
\Qqmn\ decay channels is at the level of 4-5\% and is considered 
as signal. 

In the reconstruction of the \Qqln\ events we obtain $\Cthw$ by summing 
the kinematically fitted four-momenta of the two jets. 
The decay angles of the leptonically decaying W are obtained from the 
charged lepton four-momentum, after boosting back to the parent W rest 
frame. For the hadronically decaying W it is not possible to
distinguish between the jets of the quark and anti-quark. This 
ambiguity is taken into account in the analysis described below. 

\begin{figure}[tbhp]
\begin{center}
 \epsfig{file=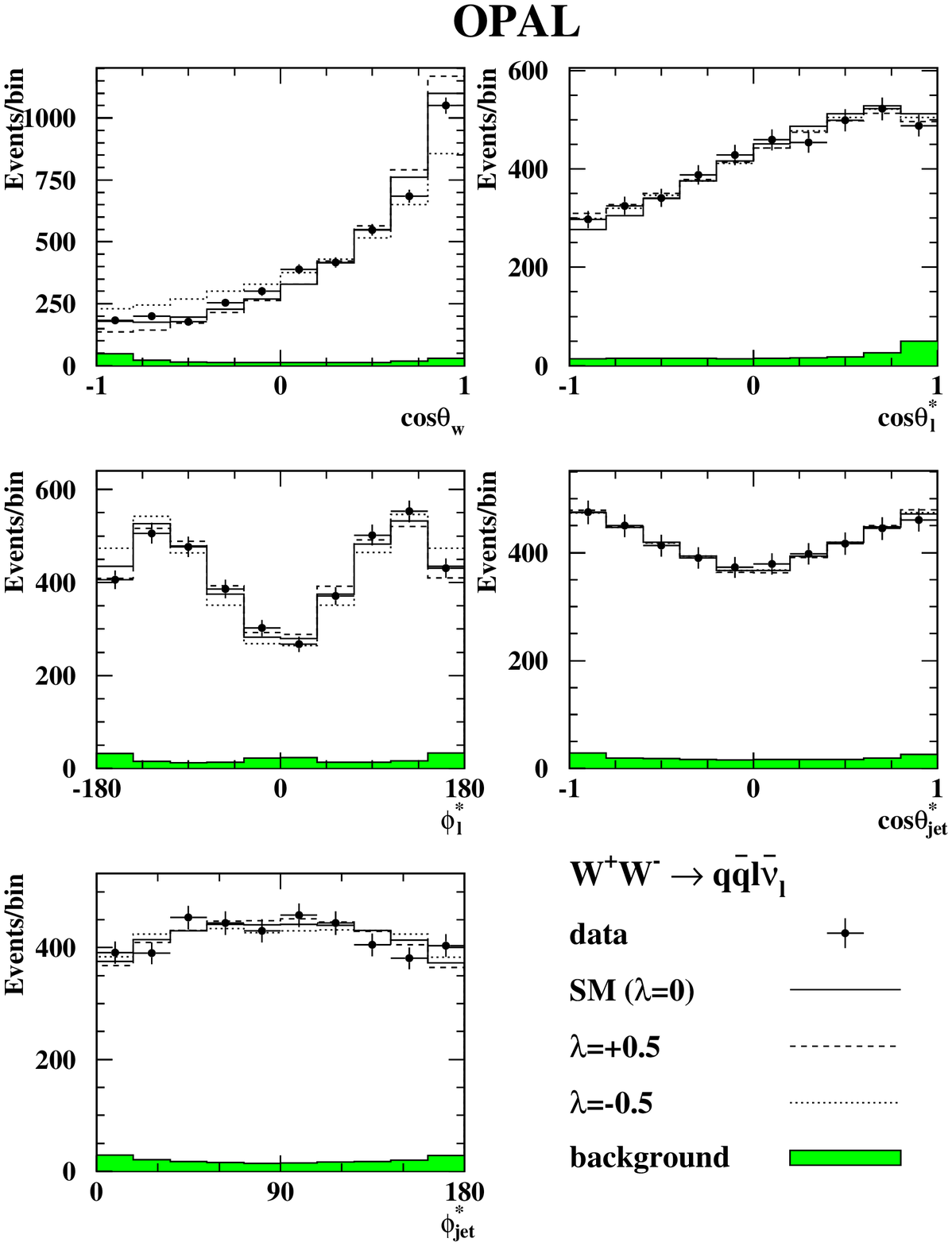,width=0.95\textwidth}
   \caption{\sl
    Distributions of the kinematic variables \Cthw, \Cthstl,
      \Phistl, \Cthstj\ and \Phistj, obtained from the \Qqln\ 
      candidates. The solid points
    represent the data. The histograms show the \MC\ expectation of the 
    \SM\ (solid line) and the cases of $\lg=+0.5$ and $\lg=-0.5$ (dashed
    and dotted lines respectively). The \MC\ histograms are normalised
    to the number of data events. The shaded histograms show the 
    non-\Qqln\ background.
    In the case of W$^+\ra\bar{\ell}\nu$ decays the value of $\Phistl$
    is shifted by 180$^{\circ}$ in order to overlay \Wp\ and \Wm\ 
    distributions in the same plot, which is valid so long as CP 
    is conserved.}
 \label{fig:angdist} 
\end{center}
\end{figure}

In Figure~\ref{fig:angdist} we show the distributions of the five
angles corresponding to the \Qqln\ data sample
and the expected distributions for $\lg = \pm 0.5$ and 0.
These expected distributions are obtained from \SM\ \MC\ samples.
To produce the $\lg = \pm 0.5$ distributions the events have been 
appropriately reweighted using the \Excalibur\ matrix element 
calculation. All \MC\ distributions are normalised to the number
of events observed in the data. The shaded histograms show the 
non-\Qqln\ background and their normalisation relative to the 
signal is according to the \SM. For the \Cthstj\ and \Phistj\
distributions the jet with $0\leq\Phistj\leq 180^{\circ}$ is
arbitrarily chosen as the quark jet from the decay of the \Wm, 
or the anti-quark jet from the decay of the \Wp.
Sensitivity to \lg\ is apparent mainly for \Cthw. The contribution
of \Cthstl, \Phistl, \Cthstj\ and \Phistj\ to the overall 
sensitivity enters mainly through their correlations with \Cthw.


\subsection{Reconstruction of \boldmath{\Qqqq} final states}
\label{subs:recqqqq}
Using the Durham $k_\perp$ algorithm~\cite{DURHAM} each selected
\Qqqq\ candidate is reconstructed as four jets whose energies are 
corrected for double counting of charged track momenta and calorimeter
energies~\cite{MT}. These four jets can be paired into W-bosons in 
three possible ways.
To improve the resolution on the jet four-momenta we perform the 
five-constraint kinematic fit B for each possible jet pairing.
In the following we consider only good jet pairings, namely 
those pairings with a successful fit B yielding a W-mass between 
70 and 90 GeV. We accept only events with at least one good jet 
pairing.

After this cut the overall efficiency is around 80\%, varying slightly
with the \Com\ energy, and the contamination from other \WW\ decays is 
below 0.3\%. The major background contribution is from \ZGqq\
where additional quarks or gluons are radiated off the primary quarks.
This contribution is between 12\% to 17\%, decreasing with \Com\ energy.
The contribution of \tgc-independent four-fermion final states 
is 4.5\%, except at 183~GeV where it drops to 3\%.
 
We use a jet charge method as an estimator for the W charge. 
For the two jets coming from one W candidate we calculate the 
momentum weighted charge of all tracks contributing to this W 
candidate as, 
  $Q_{\rm W} = \sum_{i=1}^{N_{\rm W}} q_i |p_{||,i}|^{0.5} \left/
                 \sum_{i=1}^{N_{\rm tot}} |p_{||,i}|^{0.5}\right. \/ $,
where  $N_{\rm W}$ and $N_{\rm tot}$ are the numbers of tracks in the
W candidate and in the total event respectively, $q_i$ is the charge of 
the $i^\textrm{th}$ track and $p_{||,i}$ is the momentum component 
parallel to the jet axis.
We take the W candidate with the more positive charge as the \Wp.

\begin{figure}[htbp]
  \begin{center}
\epsfig{file=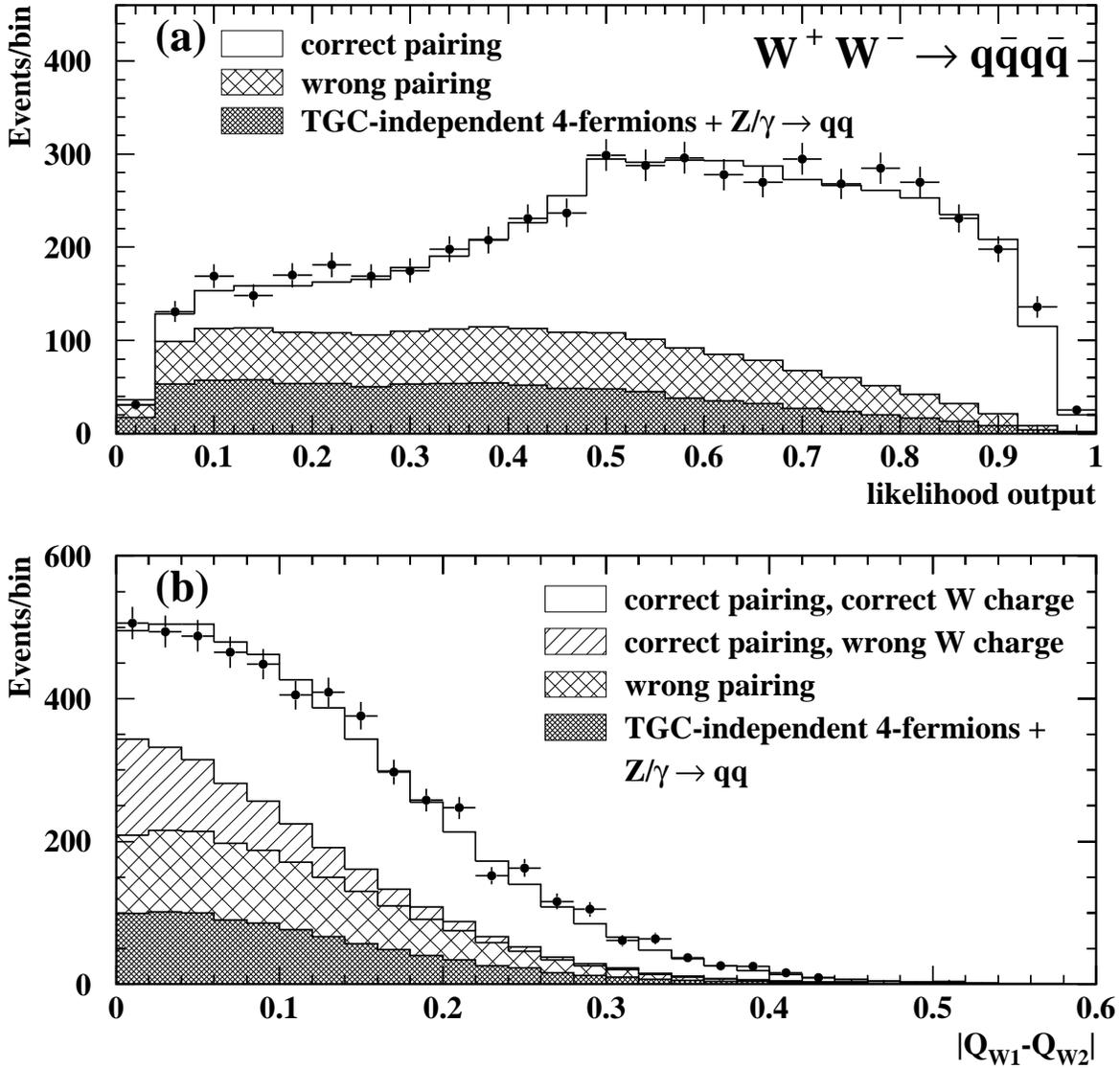,width=\linewidth,clip=}
    \caption{\sl
     Distributions for \qqqq\ data events (points) for the KandY 
     four-fermion \MC\ (histogram) of  
     (a) jet-pairing likelihood corresponding to the most likely 
     combination; 
     (b) charge separation between the two W candidates.
}   
    \label{fig:likout_qw}
  \end{center}
\end{figure}

Around 38\% of the events have more than one good jet pairing,
but the fraction of events where all three possible jet pairings
are good is below 5\%. To select the correct jet pairing, the
good pairings in the event are processed through a likelihood 
algorithm~\cite{Karlen} which takes into account correlations between
the input variables. We use as input to our likelihood the
W candidate mass as obtained from the kinematic fit B, the absolute
value of the difference between the W candidate masses as obtained 
from the kinematic fit A and the absolute value of the jet charge 
difference between the two W candidates. The algorithm has been
tuned on Monte Carlo events at each \Com\ energy
in order to give the optimal performance. The di-jet pairing with the
highest likelihood output is taken as the correct pairing.
The distribution of the highest jet pairing likelihood output 
is shown in Figure~\ref{fig:likout_qw}(a). The distribution of 
the W charge separation for all highest likelihood pairings is shown 
in Figure~\ref{fig:likout_qw}(b).

According to the \WW\
\Koralw\ Monte Carlo, the probability of selecting the correct
di-jet combination is about 80\%, and the probability of correct 
assignment of the W charge, once the correct pairing has been chosen, 
is about 77\%.
Both probabilities vary slightly with the \Com\ energy but depend
strongly on the jet pairing likelihood output.

In order to reflect this dependence on the jet pairing likelihood
output we use the output as a quality variable to classify the events
into four quality classes which are later used in the \tgc\ analysis.
The definitions of these classes and the fraction of 
accepted events in each class are listed in Table~\ref{tab:4qclasses}.
For classes with higher jet pairing likelihood output the
probability for selecting the correct di-jet combination and 
the correct W charge assignment increases. The numbers are given 
in Table~\ref{tab:4qclasses}. The population of each class is well
modelled by \MC.

\begin{table}[hbtp]
\begin{center}
\begin{tabular}{|l|c|c|c|c|} \hline
Class & Jet pairing & \% of events & Correct jet & Correct charge \\
& likelihood output & & pairing probability & assignment probability \\
\hline
a) & 0.72 - 1.00 & 27.2 & 91.8\% & 89.4\% \\ 
b) & 0.54 - 0.72 & 24.9 & 84.4\% & 74.1\% \\ 
c) & 0.35 - 0.54 & 22.4 & 75.3\% & 66.2\% \\    
d) & 0.00 - 0.35 & 25.5 & 56.4\% & 67.9\% \\    
\hline
\end{tabular}
\caption{Definition of the four quality classes of \Qqqq\ events.
The correct charge assignment probability is calculated only for those
events with correct jet pairing.} 
\label{tab:4qclasses}
\end{center}
\end{table}

\begin{figure}[htbp]
  \begin{center}
    \leavevmode

\epsfig{file=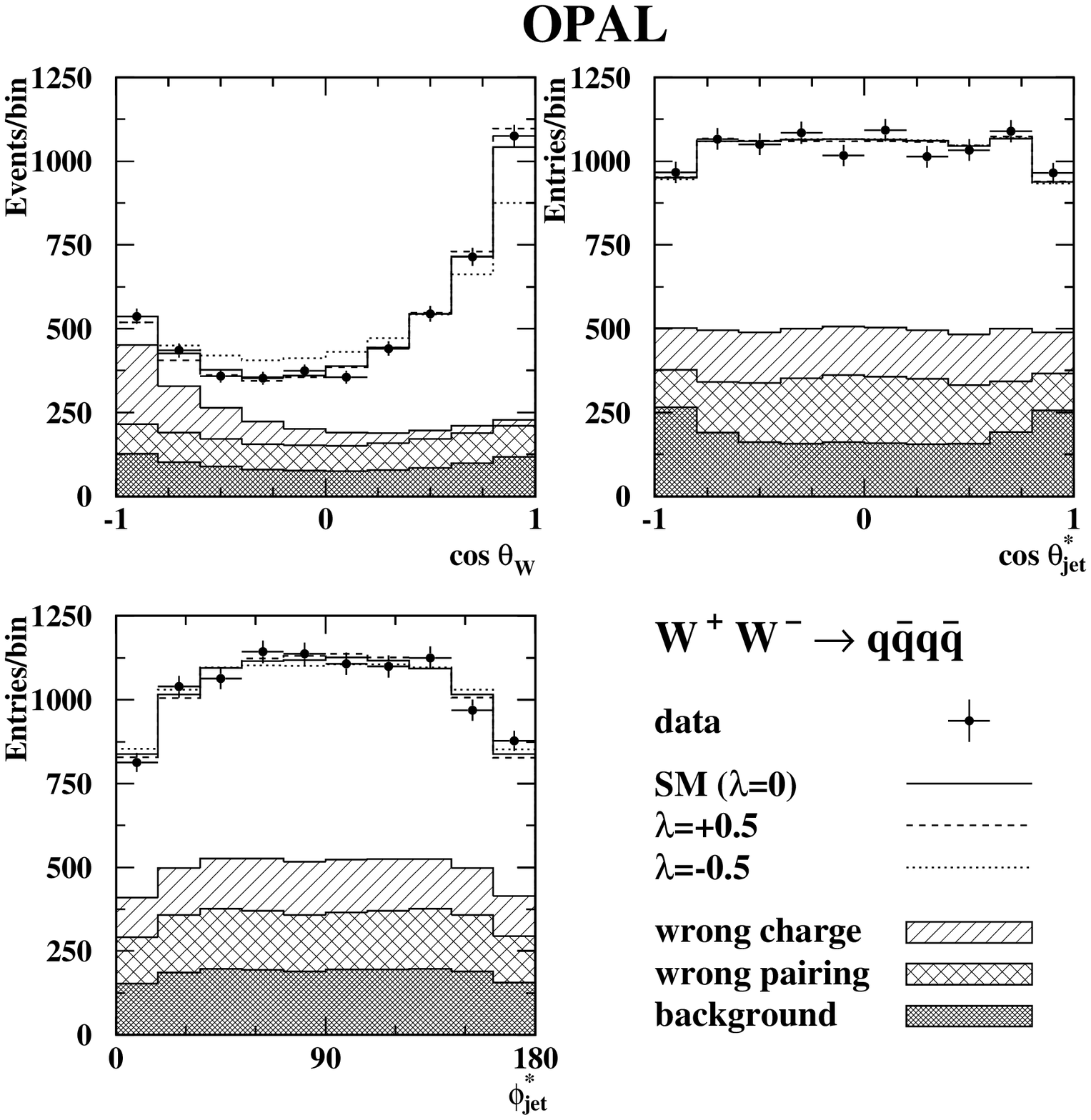,width=\linewidth,clip=}
    \caption{\sl
      Distribution of the kinematic variables \Cthw, \Cthstj, 
      and \Phistj\ for \Qqqq\ events. In the \Cthstj, and \Phistj\ distributions 
      there are two entries per event. The solid points represent 
      the data. The histograms show the \MC\ expectation of the \SM\ 
      (solid line) and the cases of \lg=+0.5
      and \lg=--0.5 (dashed and dotted lines respectively).
      The \MC\ histograms are normalised to the number of data events.
       The open histograms show the contribution from correct pairing
       and correct charge assignment. The single hatched histograms 
       represent the contribution from correct pairing, but wrong charge
       assignment. The cross-hatched histograms correspond to the 
       contribution from wrong pairing, and the dark histograms show 
       the contribution of background.} 
    \label{fig:kineqqqq}
  \end{center}
\end{figure}

The distributions of the variables \Cthw, \Cthstj\ and \Phistj\ are 
shown in Figure~\ref{fig:kineqqqq}, along with the expected 
distributions for the \SM\ (\lg=0) and for \lg=$\pm0.5$. In the
\Cthstj\ and \Phistj\ distributions there are two entries per event,
and according to our convention (see Section~\ref{subs:recqqln}),
the jet with $0\le\Phistj\le 180^{\circ}$ is chosen as the quark jet 
from the decay of the \Wm, or the anti-quark jet from the decay of 
the \Wp. The \MC\ histograms are normalised to the number of data 
events. The shaded histograms represent the separate contributions 
from wrong pairings, correct pairing but wrong charge and background. 
Their relative normalisation is according to the \SM. 
The data are again described well by the \SM\ prediction.


\subsection{Reconstruction of \boldmath{\Lnln} final states}
\label{subs:reclnln}
 
The reconstruction of \Lnln\ final states is 
possible, using kinematic fit C, only if there is no $\tau$-lepton 
in the final state. To reject events with $\tau$-leptons we use the lepton
identification algorithm designed for the W branching ratio
measurement~\cite{sigww189}.
The event rate analysis selection sometimes selects events with only one
high momentum charged lepton reconstructed in the detector: such events
were removed from the angular distribution analysis as the momenta of both
charged leptons must be measured. As a result of these two cuts, the 
efficiency for \Lnln\ events with $\ell,\ell^{\prime}={\rm e},\mu$ 
drops from $\sim$88\% to $\sim$73\%, varying slightly with \Com\ energy.
The contamination left in the sample from \Lnln\ final states with a 
$\tau$-lepton is 14\% and the background from other final states is 2\%.

Kinematic fit C has zero constraints and reduces to solving a 
quadratic equation. In the ideal case of no measurement errors and 
satisfying the conditions 
where fit C is valid, namely no ISR and narrow W width, one expects 
to obtain two real solutions corresponding to a two-fold ambiguity 
in the angle set $\cos\theta_W$, $\phi_1^*$ and $\phi_2^*$. There is
no ambiguity in the angles $\theta_1^*$ and $\theta_2^*$ which have a
one-to-one correspondence with the lepton momenta. 
In realistic conditions, however, including ISR, finite W width and 
measurement errors, one may obtain no real solution, but a pair 
of complex conjugate solutions. The fraction of events with
complex solutions falls with increasing \Com\ energy from 30\% to 24\%,
in agreement with \MC\ prediction. In these events, the imaginary 
parts of the complex solutions are set to zero, yielding a single real
angle set. The total weight of all events is the same, independent of
whether one or two solutions are employed. In the latter case, the two
solutions each are assigned a weight of one half.

\begin{figure}[htbp]
\epsfig{file=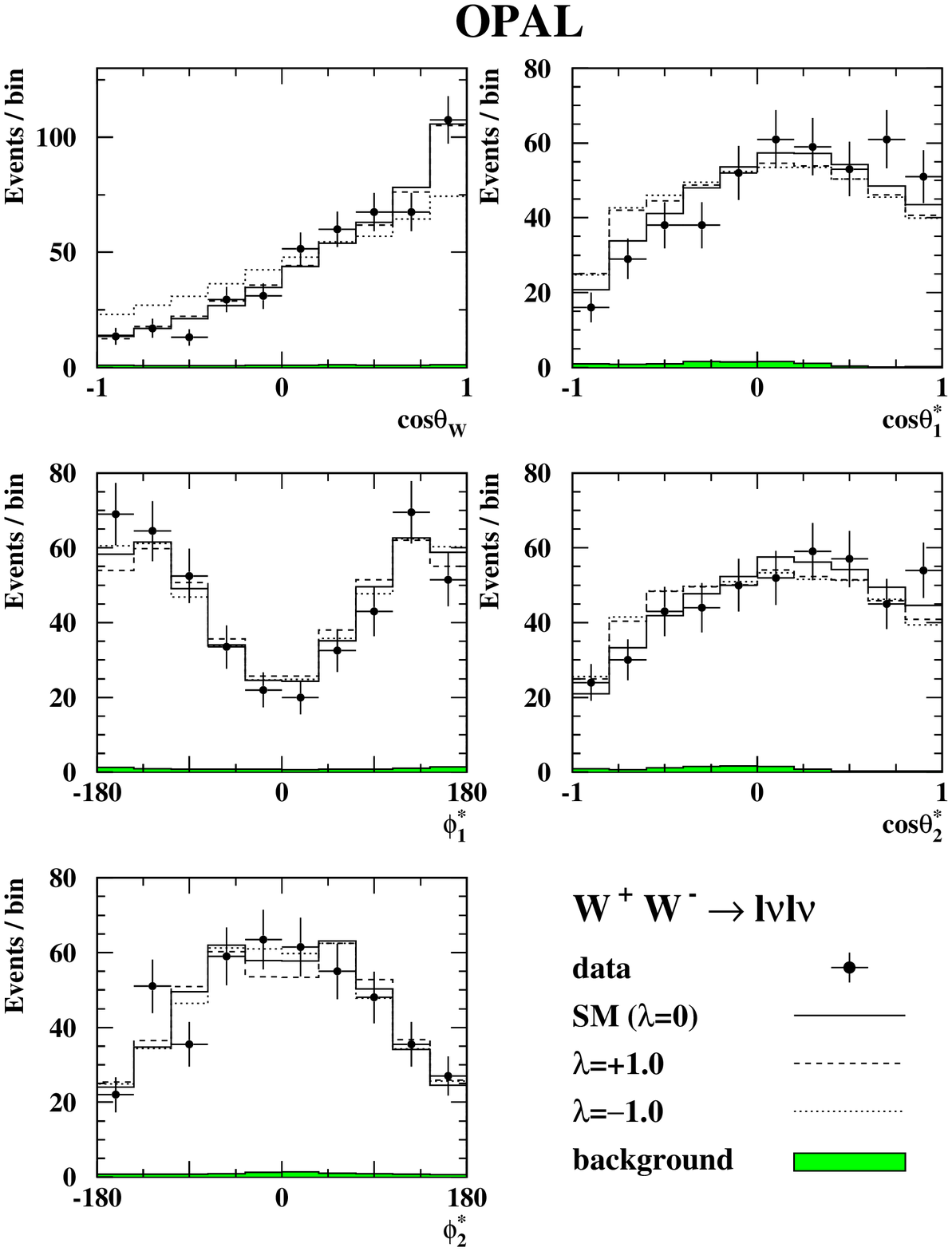,bbllx=5pt,bblly=17pt,bburx=515pt,bbury=690pt,
     width=0.92\textwidth,clip=}
\caption{\sl
Distributions of the kinematic variables \Cthw, \Cthsto, \Phisto, \Cthstt\
and \Phistt\ in the \Lnln\ analysis,
where \Cthsto, \Phisto\ (\Cthstt, \Phistt)
are the decay angles of the charged lepton (anti-lepton) from the \Wm\ 
(\Wp) in the parent W rest frame. All events enter
with a total weight of one, but in the case of events with two ambiguous
solutions, each solution enters with a weight of 0.5.
The histograms, normalized to the number of data events, show the 
expectation of the \SM\ and the cases of $\lg=\pm 1$. 
The shaded histograms represent the contribution of the background.
}
\label{fig:dacostw}
\end{figure}

Figure~\ref{fig:dacostw} shows the distributions of the five reconstructed
angles and the expected distributions for \lg=0, $\pm 1$. The \MC\ 
histograms are normalized to the number of data events. The shaded 
histograms represent the contribution of background and
their normalisation relative to the signal is according to the \SM.
The data distributions agree with the expectation of the \SM.


\section{Optimal observable analysis}  
\label{sec:oo} 

The angular distributions from all final states are consistent with the 
\SM\ expectation and no evidence is seen for any significant
contributions from anomalous couplings. For a quantitative study we
use the method of optimal observables in the same way as 
in~\cite{tgc189-analysis}. This method relies on the linear dependence
of the triple gauge vertex Lagrangian on the \tgc s, corresponding
to a second-order polynomial dependence of 
the differential \xse, 
\[
\frac{d\sigma(\Omega,\al)}{d\Omega} = 
  S^{(0)}(\Omega) + \sum_i \al_i \cdot S_i^{(1)}(\Omega)  + 
\sum_{i,j} \al_i \al_j \cdot S^{(2)}_{ij}(\Omega),\ \ \
   \al_i \ = \ \dkg,\ \dgz,\ \lg \  {\rm and} \  \gfz,  
\]
where $S^{(0)}$, $S_i^{(1)}$ and $S^{(2)}_{ij}$ are functions of the 
five phase-space variables, 
$\Omega$=(\Cthw, \Cthsto, \Phisto, \Cthstt, \Phistt).
For this kind of dependence it has been shown~\cite{Fanourakis} that
all information contained in the phase-space variables $\Omega$, 
which is relevant to the four couplings, is retained 
in the whole set of 14 observables 
$\OO_i^{(1)}=S_i^{(1)}(\Omega)/S^{(0)}(\Omega)$,
$\OO_i^{(2)}=S_{ii}^{(2)}(\Omega)/S^{(0)}(\Omega)$ and
$\OO_{ij}^{(2)}=\OO_{ji}^{(2)}=S_{ij}^{(2)}(\Omega)/S^{(0)}(\Omega)$ 
$(i,j=1,2,3,4)$. In this analysis we use the mean values of  
$\OO_i^{(2)}$ and $\OO_{ij}^{(2)}$ in addition to the mean values 
of $\OO_i^{(1)}$~\cite{Diehl}.

In the one-parameter fit of $\al_i$ we only use 
the corresponding first order, $\OO_i^{(1)}$, and second order,
$\OO_i^{(2)}$, observables. In this way, the other couplings are not
used in the fit. Similarly, in two-parameter fits where two 
couplings are allowed to vary simultaneously, we use
only the five observables corresponding to these
couplings and their mutual correlations. When three couplings are
allowed to vary, nine observables are used. 

The calculation of the mean optimal observables and their 
expected values is done separately for each centre-of-mass energy 
and for each of the \Qqln, \Qqqq\ and \Lnln\ final states. 
The optimal observables are constructed for each event $k$ with the 
set of phase-space variables $\Omega_k$ using the analytic expression
for the CC03 Born differential \xse\ to calculate the values 
of $S^{(0)}(\Omega_k), \ S_i^{(1)}(\Omega_k), \ S_{ii}^{(2)}(\Omega_k)$ 
and $S_{ij}^{(2)}(\Omega_k)$. This calculation 
takes into account the reconstruction ambiguities for each particular
final state. For the \Qqqq\ channel, the ambiguity in the W-charge 
determination is taken into account by weighting the 
$S^{(0)}$, $S_i^{(1)}$ and $S^{(2)}_{ij}$ functions for each
charge hypothesis with the corresponding probability. This 
probability is determined from the Monte Carlo as a function
of the estimated charge difference $|Q_{W_1}-Q_{W_2}|$
between the two W candidates in the event. 

For \Qqln\ and \Qqqq\ events the mean optimal observables are 
calculated using weights assigned to the event according to its 
quality class (see Tables~\ref{tab:classes} and \ref{tab:4qclasses}).
For \Qqln\ events, the weight for a particular class is inversely
proportional to the variance of the optimal observable 
distribution of the events in that class. For \Qqqq\ events, the
weights are calculated in a similar way, using the variances of the
optimal observable {\em resolution} distributions. In this way,
events where the quality of the reconstruction is poor obtain 
a lower weight. Using this weighted mean has been found, as 
expected, to enhance our sensitivity.  

The expected mean values of each observable as a function 
of the \tgc s are calculated from four-fermion reference \MC\ samples
generated according to the \SM.
The \Excalibur\ matrix element calculation is used 
to reweight the \MC\ events to any \tgc\ value required. The contribution 
of background events is also taken into account in the calculation of 
the expected mean values using corresponding \MC\ samples.

A $\chi^2$ fit of the measured mean values \MOO\ to the corresponding 
expectations \OOEXP{}{}(\al) is performed to extract the couplings 
from the data. The covariance matrix for the mean values is calculated 
from the \MC\ events and scaled to the number of data events.

In some of our fits the $\chi^2$ function has a double minimum and 
therefore it is necessary to validate the errors obtained from 
the $\chi^2$ fits. This is performed with a large number of Monte Carlo 
``experiments''. In each of these ``experiments'', we use for each 
centre-of-mass energy and each final state a subsample of \SM\ \MC\ 
events. The size of each subsample corresponds to the luminosity for 
that final state at the particular \Com\ energy. Background events 
are also included in these subsamples
in the appropriate proportions. For each ``experiment'', the various
subsamples are analysed in the same way as the real data events.
The distributions of the fit results are centred around the expected
values, but there are some non-Gaussian tails.
We therefore test the reliability of the error interval, as given by
the region where $\chi^2$ is no more than 1 above its minimum value, 
by counting the fraction of ``experiments'' where the correct value
falls within this region. 
The error estimate is considered to be reliable
if the calculated fraction is consistent with 68\%; otherwise the
corresponding elements of the covariance matrix for the \MOO\ values 
are scaled up by the necessary factor to obtain 68\% of the subsamples 
within the error interval. The resulting scaled covariance matrix 
is also used to analyse the real data. This is done 
separately for each fit. The resulting scale factors vary 
between 1 and 1.076 for the one-parameter fits. However, in the 
three-parameter fit of \dkg, \dgz\ and \lg, where double minima occur
more frequently, a larger scale factor of 1.28 is obtained.
These scale factors have been found to be appropriate also for the
95\% C.L. intervals.


\section{Systematic errors in the angular distribution analysis}
\label{sec:sys}

The following sources of systematic uncertainty are considered, as 
listed in Table~\ref{tab:sys}.
\begin{itemize}
\item[a)]There is a theoretical uncertainty in the \OO(\al) 
  correction due to missing higher orders in
  the \YFSww\ \MC\ generator program. 
  As a conservative estimate of this uncertainty, this \OO(\al)
  correction is removed, degrading the events to \Koralw1.42 level. 
  The efffect of this change is taken as a systematic uncertainty. 
\item[b)]\Koralw\ includes ISR up to $\OO(\al^3)$. As a conservative
  estimate of missing higher orders, we replace it with ISR up
  to $\OO(\al)$ only and take the difference as a systematic 
  uncertainty.
\item[c)]Another uncertainty in the signal \MC\ generator is
  related to the \Pythia\ fragmentation and hadronisation model which 
  is used. To study the effect of this uncertainty we compare
  results of the analysis when applied to 
  \WW\ Monte Carlo samples generated with \Koralw1.42\ using either 
  \Pythia, \Herwig\ or \Ariadne\ for the fragmentation and 
  hadronisation phase. The effect of changing to \Herwig\ is the
  larger and is used to estimate the uncertainty.
  The main effect is on the \Qqqq\ final state, where \Herwig\ predicts
  a higher probability of correct jet pairing and correct W charge
  assignment than \Pythia. This effect is partially due to the average
  charged multiplicity for light quarks being lower in our tuned version 
  of \Herwig\ than in \Pythia. Since the data are in agreement with
  \Pythia~\cite{MULTIP}, we weight the \Herwig\ Monte Carlo events so 
  as to reproduce the same charged multiplicity distribution as 
  \Pythia. The effect of this weighting is a reduction of the 
  systematic error by about 15-30\%.
\item[d)] To assess the uncertainty in the \tgc\ matrix element
  calculation which is used to weight our Monte Carlo events,
  we replace the calculation from the \Excalibur\ \MC\ 
  generator program with the one from \Grace.  
\item[e)] The \MC\ simulation of the OPAL detector has been 
  studied using back-to-back jets and leptons in \Zz\ events 
  collected during calibration runs which were taken regularly
  every year between the high energy runs.  
  Some scaling and smearing had to be applied to the reconstructed 
  jets and leptons in the \MC\ in order to achieve the best
  agreement with data. The uncertainties in these corrections
  correspond to variations in our results which are taken as
  systematic errors. These corrections are usually applied
  {\em after} the event selection. To estimate their effect on the
  selection of our sample, they were applied as a systematic check
  {\em before} the event selection, and the difference from the 
  standard result was added in quadrature with the effects of the
  uncertainties on the corrections.
\item[f)]Uncertainties in the background estimation
  are determined by varying both its shape and normalisation.
  To study the effect of the shape we use alternative samples for 
  each background source, keeping the normalisation fixed.
  For the \tgc-independent four-fermion background we use 
  \Excalibur\ and \Grace\ instead of \Koralw. 
  For the \ZGqq\ background we use \KK\ with \Herwig\ or
  \Ariadne\ rather than \Pythia\ fragmentation. Finally, the 
  \Herwig\ two-photon samples are replaced with \Ftogen. For the 
  background normalisation uncertainty the contribution from
  each source is scaled, one at a time, by the factors 1.2, 
  1.3 and 2.0 respectively. These scale factors
  are determined from studies with control data samples and from
  the differences between the alternative \MC\ generators, as
  explained in~\cite{sigww189}. 
\item[g)]The consequences of the uncertainties in the \Com\ energy
  and the W mass are estimated as follows. For the \Com\ energy 
  uncertainty of 40~MeV, \MC\ samples at different \Com\ 
  energies are used as pseudo-data. For the W mass  
  we assume \Mw=80.456$\pm$0.059 \GeVcc~\cite{MWTEV} 
  which is 0.126~\GeVcc\ higher than the one used in our
  \MC\ generators. We do not correct our \tgc\ fit result for this 
  shift in \Mw, but it is accounted for in our systematic error,
  along with the \Mw\ measurement error. The combined effect is 
  assessed using \MC\ samples generated with different W masses.
\item[h)]Bose-Einstein correlations (BEC) in \Qqqq\ events might affect
  the reconstruction of the W bosons and measured W charge distribution.
  We investigate this effect with a Monte Carlo program simulating BEC
  via re-adjustment of final state momenta using the LUBOEI
  model~\cite{BEC}. When both inter-W and intra-W correlations are
  taken into account, the bias is found to be larger than the case
  where only intra-W boson correlations are present.
\item[i)]The jet reconstruction and the measured W charge distribution 
  in \Qqqq\ events might also be affected by colour reconnection. 
  This effect is investigated with several MC samples generated according 
  to the Sj\"ostrand-Khoze \cite{SK} models. We consider model I 
  with colour reconnection probabilities of 34\% and 97\%. 
  Models II and II' are also considered. Model I with colour
  reconnection probability of 97\% produces the largest bias.
\end{itemize}

The \MC\ samples are larger than the data by a factor of 150 - 900,
depending on the \Com\ energy. Therefore, the uncertainty due to 
\MC\ statistics is neglected.

The effect of each of these sources on the mean values of the optimal
observables is estimated separately for each centre-of-mass energy and
for each final state. The
deviations obtained in the mean observables are used to construct a
systematic error covariance matrix, where the systematic deviations 
from each source are assumed to be fully correlated between all 
centre-of-mass energies. For the combination of all final states, 
the covariance matrix is extended to incorporate the three final 
states. All uncertainties, except for those associated with the 
background, are assumed
to be fully correlated between the relevant final states. 
This matrix is added to the statistical covariance matrix,
and the fit is redone to obtain the \tgc\ parameters with their total 
(statistical and systematic) errors. 

\begin{table}[bhtp]
\begin{center}
\begin{tabular}{|ll|c|c|c|c|}\hline 
  \multicolumn{2}{|c|}{Source}&  \dkg & \dgz & \lg & \gfz \\ \hline
 a) & \OO(\al) correction &\ddkgOA &\ddgzOA &\dlamOA &\dgfzOA \\
 b) & ISR                 &\ddkgIS &\ddgzIS &\dlamIS &\dgfzIS \\
 c) & Fragmentation       &\ddkgFR &\ddgzFR &\dlamFR &\dgfzFR \\
 d) & \tgc\ matrix element &\ddkgTG &\ddgzTG &\dlamTG &\dgfzTG \\
 e) & Detector simulation &\ddkgDE &\ddgzDE &\dlamDE &\dgfzDE \\
 f) & Background          &\ddkgBG &\ddgzBG &\dlamBG &\dgfzBG \\
 g) & \roots, \Mw         &\ddkgEB &\ddgzEB &\dlamEB &\dgfzEB \\ 
 h) & Bose-Einstein correlations &\ddkgBE &\ddgzBE &\dlamBE &\dgfzBE \\ 
 i) & Colour reconnection &\ddkgCR &\ddgzCR &\dlamCR &\dgfzCR \\ \hline
    & Total               &\ddkgTO &\ddgzTO &\dlamTO &\dgfzTO \\ \hline
\end{tabular}
\end{center}
\caption{ Contributions to the systematic uncertainties in the 
  determination of the \tgc\ parameters. The total systematic errors
  are obtained by combining all contributions at the optimal 
  observable level and then calculating the effect on the \tgc\
  parameters.}
\label{tab:sys}
\end{table}

For demonstration, we also calculate the separate contribution from 
each source of systematic error to the error on the \tgc\ parameters,
after combining the different \Com\ energies and final states.
These are listed in Table~\ref{tab:sys}.


\section{Event rate TGC analysis}
\label{sec:rate}

The event rate study, unlike the angular distribution analysis, 
requires a detailed investigation of the systematic uncertainties
associated with the overall selection efficiency. This 
investigation is part of the W-pair \xse\ analysis and has to be 
done separately for each \Com\ energy and for each final state.
So far, it has been completed for \Com\ energies of 
183~GeV~\cite{tgc183-analysis} and 189~GeV~\cite{sigww189}. 
The \tgc\ event rate analysis is therefore also restricted 
to these two \Com\ energies. In the case where the angular distribution
analysis has two \tgc\ solutions, one being far away from the \SM\
value, the results from this restricted event rate analysis would still
be sufficient to reject that solution. The data from higher \Com\ energies 
would make a negligible additional contribution to our results. 
Compared with the angular distributions, the sensitivity of the 
event rates to \tgc s around the \SM\ values is much smaller and 
its dependence on the overall luminosity is weaker.

The event rate analysis has been described already in our previous 
publications~\cite{tgc183-analysis,tgc189-analysis}. The numbers
of expected events are re-evaluated for this analysis using 
the KandY \MC\ samples to obtain the \SM\ expected number of events.

The same selection, efficiency and background 
estimates as in the total \xse\ analysis are used here.
The numbers of selected events in each final state and for each of the 
two \Com\ energies are quoted in Table~\ref{tab:rate} along with 
the \MC\ expectations. The systematic errors on the expected
values are due to \MC\ statistics, 0.5\% uncertainty 
in the total cross-section, 0.2\% uncertainty in the luminosity,
uncertainties due to data/MC differences, tracking losses, detector
occupancy and fragmentation. A detailed description of
all these sources can be found in~\cite{sigww189}.
The expected numbers of signal events include contributions from 
all four-fermion final states which can be produced by diagrams
involving the WW$\gamma$ and WWZ triple gauge boson vertices.
The other four-fermion final states are considered as 
background along with the two-fermion final states and   
two-photon processes.
 
\begin{table}[htbp]
 \begin{center}
 \begin{tabular}{||c|c|c||c||c|c|c||} \hline
\roots\ & Luminosity & Final & Observed  &
  \multicolumn{3}{|c||}{Expected events} \\ \cline{5-7}
(GeV) & (\Ipb) & state & events & Total & Signal & Background \\ \hline
182.7 & $57.21\pm0.25$ & \lnu\lnubar &  78 &$77.5\pm3.1$& $76.1\pm2.2$&
    $1.4\pm2.2$ \\
      &                & \Wqq\lnu    & 361 & $364\pm 6$ &  $337\pm 6$ &
     $27\pm 3$  \\
      &                & \Wqq\Wqq    & 438 & $436\pm10$ &  $344\pm 5$ &
     $92\pm 9$  \\  \hline
188.6 &$183.05\pm0.40$ & \lnu\lnubar &276 &$287.2\pm5.0$&$272.6\pm4.1$&
   $14.6\pm2.9$ \\  
      &                & \Wqq\lnu    &1246 &$1248\pm16$ & $1171\pm13$ &
     $76\pm10$  \\
      &                & \Wqq\Wqq    &1546 &$1497\pm26$ & $1189\pm14$ &
    $309\pm21$  \\  \hline
 \end{tabular}
\caption{Observed and expected numbers of events in each \WW\ final
state for the \Com\ energies of 182.7 GeV and 188.6 GeV. 
The separation between signal and background is explained in the text.
All the quoted errors are systematic.} 
\label{tab:rate}
 \end{center}
\end{table}

The total numbers of expected events are consistent within the 
statistical and systematic errors with the observed numbers.
Therefore, there is no evidence for any significant contribution
from anomalous couplings. 
A quantitative study of \tgc s from these results is performed by 
comparing the numbers of observed events in each of the three event
selection channels at each of the two \Com\ energies with the 
expected numbers of signal and background events, where the number
of signal events is parametrised as a second-order polynomial in 
the \tgc s. This parametrisation is based on the 
second-order polynomial dependence of the \xse, as was used 
also in the optimal observable analysis (see 
Section~\ref{sec:oo}). The polynomial coefficients in terms 
of event rate for 
each final state are calculated by weighting our \MC\ events to 
correspond to various \tgc\ values using a four-fermion matrix 
element calculation procedure taken from the \Excalibur\ \MC\ 
generator program.  

The probability to observe the measured number of candidates, given the 
expected value, is calculated using a Gaussian distribution where
the errors also include the systematic uncertainties.
We assume full correlation between the systematic uncertainties
for the three final states except for efficiency and 
background which are taken to be uncorrelated. All systematic  
uncertainties are assumed to be fully correlated between the two 
different \Com\ energies.
 

\section{Combined TGC results}
\label{sec:combtgc}

The results of the various fits are listed in Table~\ref{tab:res}.
We present first the results of the angular distribution analysis
without systematic errors obtained by performing one-parameter fits
for each final state simultaneously for the eight \Com\ energies.
For comparison, the average fit errors of the \MC\ ``experiments'' 
are also quoted in Table~\ref{tab:res}, demonstrating the different 
sensitivities for the various final states and \tgc\ parameters.
In most cases, the fit errors are rather close to the expected ones,
except for \dkg, where the fit errors for some final states are also
very asymmetric due to the proximity of a second minimum in the $\chi^2$ 
function. 

\begin{figure}[htbp]
\epsfig{file=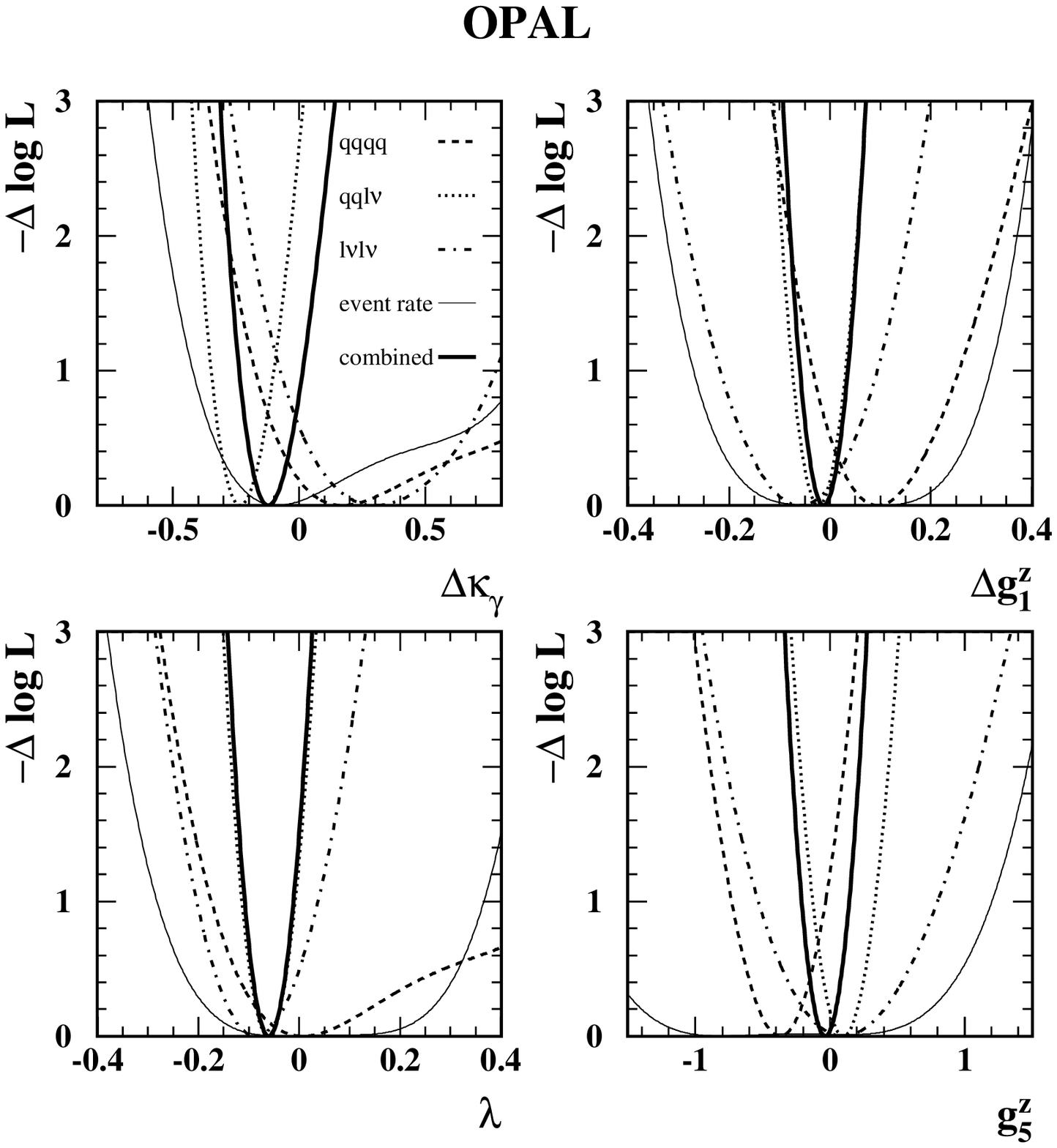,bbllx=55pt,bblly=180pt,bburx=535pt,bbury=687pt,
     width=\textwidth,clip=}
\caption{\sl
Negative log-likelihood curves obtained from the 
angular distribution information of W-pair events taken at the \Com\
energies of 183 -- 209 GeV, separately for the three different final  
states: \Qqln\ (dotted lines), \Qqqq\ (dashed lines) and \Lnln\
(dash-dotted lines). The thin solid lines describe the contribution
of the event rate information from all final states corresponding to
\Com\ energies of 183 and 189 GeV. The systematic errors are included. 
The thick solid lines are obtained by combining all sources of 
information. The curves for each \tgc\ parameter are obtained by
setting the other three parameters to their \SM\ values. 
}
\label{fig:channel}
\end{figure}

Next we present the results of the angular distribution
analysis with the systematic errors included as explained in 
Section~\ref{sec:sys}. This modifies the relative contributions to 
the fit of the different optimal observables and different \Com\ 
energies, as well as the correlations between them. Therefore, 
not only the fit errors are affected, but also the fit results 
themselves. The corresponding \LL\ curves\footnote{
Following the convention in LEP, we prefer to display our results
by the logarithm of the likelihood function, \LL, related to
the $\chi^2$ function by, --\LL= $\chi^2/2$. The plotted function
is -$\Delta$\LL, which is obtained by subtracting from --\LL\ its
minimum value.}
for the different channels 
are shown in Figure~\ref{fig:channel}. The results from the
three final states are combined, taking into account the 
common systematic errors as explained in Section~\ref{sec:sys},
and the combined results are listed in Table~\ref{tab:res}.
  
\renewcommand{\arraystretch}{1.2}
\begin{table}[bhtp]
\begin{center}
\begin{tabular}{|l|c|c|c|c|} \hline 
                      & \dkg    & \dgz    &  \lg    &  \gfz   \\ \hline 
\underline{\Qqln}     &         &         &         &         \\
Without systematics   & \qlkgst & \qldgst & \qllast & \qlgfst \\
Expected errors       & \qlkgex & \qldgex & \qllaex & \qlgfex \\
Including systematics & \qlkgsy & \qldgsy & \qllasy & \qlgfsy \\
Fit $\chi^2$/d.o.f.   & \qlkgcs & \qldgcs & \qllacs & \qlgfcs \\ \hline
\underline{\Qqqq}     &         &         &         &         \\
Without systematics   & \qqkgst & \qqdgst & \qqlast & \qqgfst \\
Expected errors       & \qqkgex & \qqdgex & \qqlaex & \qqgfex \\
Including systematics & \qqkgsy & \qqdgsy & \qqlasy & \qqgfsy \\
Fit $\chi^2$/d.o.f.   & \qqkgcs & \qqdgcs & \qqlacs & \qqgfcs \\ \hline
\underline{\Lnln}     &         &         &         &         \\
Without systematics   & \lvkgst & \lvdgst & \lvlast & \lvgfst \\
Expected errors       & \lvkgex & \lvdgex & \lvlaex & \lvgfex \\
Including systematics & \lvkgsy & \lvdgsy & \lvlasy & \lvgfsy \\
Fit $\chi^2$/d.o.f.   & \lvkgcs & \lvdgcs & \lvlacs & \lvgfcs \\ \hline
\underline{\rm All WW final states}  &  &    &         &         \\
Including systematics & \shadkg & \shadgz & \shalam & \shagfz \\ 
Including event rate  & \comdkg & \comdgz & \comlam & \comgfz \\ 
Fit $\chi^2$/d.o.f.   & \cokgcs & \codgcs & \colacs & \cogfcs \\ 
\SM\ $\chi^2$/d.o.f.  & \smkgcs & \smdgcs & \smlacs & \smgfcs \\ 
95\% C.L. limits      & \dkgint & \dgzint & \lamint & \gfzint \\ 
\hline 
\end{tabular}
\caption{Measured values of the \tgc\ parameters from the angular
distributions of each WW final state, using \Com\ energies of 
183 -- 209 GeV
and the results after combining all final states. For the individual 
final states we also list the results before including the systematic
errors and the expected statistical errors. We also list the results
after combining with the W-pair event rate information from \Com\
energies of 183 and 189 GeV. 
The quality of the fits is demonstrated by 
their respective $\chi^2$ values and numbers of degrees of 
freedom. The compatibility with the \SM\ is shown by the 
$\chi^2$ values obtained when the \tgc s assume their \SM\ values.
}
\label{tab:res}
\end{center}
\end{table}
\renewcommand{\arraystretch}{1.0}

The results of the event rate analysis are presented by \LL\ curves
shown by the thin solid lines in Figure~\ref{fig:channel}. They are
combined with the results of the angular distribution analysis
and the combined \LL\ plots are shown
by the thick solid lines. The correlation
between the results of the two analyses, which is due to 
uncertainties in the background normalisation, \Com\ energy and 
W mass, is below 7\%. Therefore this correlation is neglected in the 
combination. The numerical results of the combination are listed in 
Table~\ref{tab:res} along with the corresponding $\chi^2$/d.o.f.
As expected, the event rate information has relatively little 
impact on the final results. The 95\% confidence level intervals
are also listed in Table~\ref{tab:res}.
To show the compatibility with the \SM\ we also list in the table
the $\chi^2$/d.o.f. for \SM\ \tgc\ values.

\begin{figure}[htbp]
\epsfig{file=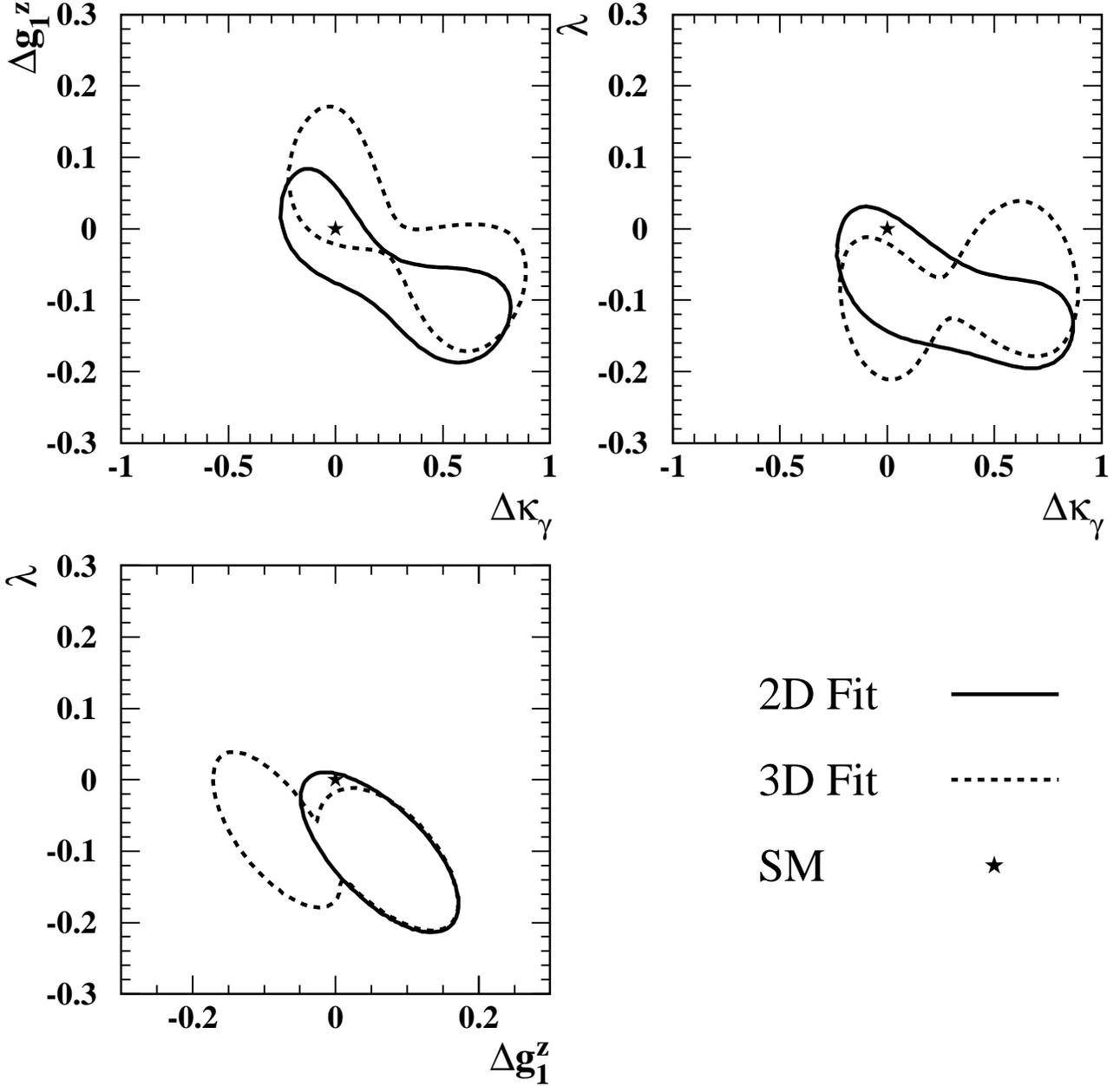,bbllx=20pt,bblly=160pt,bburx=525pt,bbury=687pt,
     width=\textwidth,clip=}
\caption{\sl
The two-dimensional correlation contours corresponding 
to $-\Delta\LL=2.995$ for different pairs of \tgc\ parameters,
as obtained from two-parameter (solid curves) and three-parameter
(dashed curves) fits.  The value 2.995 corresponds to a 95\% C.L. 
region in two dimensions.
In the case of the three-parameter fit, the curve 
is a projection onto the two-dimensional plane of the envelope 
of the three-dimensional $-\Delta\LL=2.995$ surface.
The stars indicate the \SM\ expectations.}
\label{fig:2d3d}
\end{figure}

To study correlations between the C- and P-conserving \tgc\
parameters, \dkg, \dgz\ and \lg, we perform 
fits where two parameters are allowed to vary while the third 
parameter and \gfz\ assume their \SM\ values and 
are not used in the fit, as explained in Section~\ref{sec:oo}. 
Similarly, we also perform a three-parameter fit where all 
three parameters are allowed to vary, leaving out only \gfz.
Figure~\ref{fig:2d3d} shows the 95\% C.L. contour plots obtained from 
the two-parameter fits and the corresponding two-dimensional
projections of the three-parameter fit. 
The contour plots show a double minimum structure. Adding information 
from the single W channel, where the --\LL\ function tends to have a 
steep rise with increasing \dkg~\cite{OTHERSW}, is expected to
resolve this ambiguity.

A $\chi^2$ test can be used to study the compatibility of our data 
with the \SM, in the same way as was done for the one parameter fits.
Inserting the \SM\ values for the \tgc\ parameters we obtain 
$\chi^2=211.6$ for 222 degrees of freedom.

\section{Summary}
\label{sec:summary}  

Using a sample of \WW\ candidates collected 
at \LepII\ at \Com\ energies of 183 -- 209~\GeV, we measure the 
CP-conserving triple gauge boson couplings \kg, \gz, \lg\ and \gfz. 
For this measurement we use both the angular information of W-pair
events and their total event rate. The results 
obtained are: 
\begin{eqnarray*}
\kg & = & \reskg, \\     
\gz & = & \resgz, \\    
\lg & = & \reslam, \\
\gfz & = & \resgfz,   
\end{eqnarray*}
where each parameter is determined from a single-parameter fit,
while the other three parameters assume their \SM\ values. 
These results supersede those from our previous 
publications~\cite{tgc161-analysis,tgc172-analysis,tgc183-analysis,
tgc189-analysis}. They are all
consistent with the \SM\ predictions of 1, 1, 0 and 0 respectively.
This measurement constitutes strong evidence for the gauge
structure of the \SM. 


\section*{Acknowledgements:}

We particularly wish to thank the SL Division for the efficient operation
of the LEP accelerator at all energies
 and for their close cooperation with
our experimental group.  In addition to the support staff at our own
institutions we are pleased to acknowledge the  \\
Department of Energy, USA, \\
National Science Foundation, USA, \\
Particle Physics and Astronomy Research Council, UK, \\
Natural Sciences and Engineering Research Council, Canada, \\
Israel Science Foundation, administered by the Israel
Academy of Science and Humanities, \\
Benoziyo Center for High Energy Physics,\\
Japanese Ministry of Education, Culture, Sports, Science and
Technology (MEXT) and a grant under the MEXT International
Science Research Program,\\
Japanese Society for the Promotion of Science (JSPS),\\
German Israeli Bi-national Science Foundation (GIF), \\
Bundesministerium f\"ur Bildung und Forschung, Germany, \\
National Research Council of Canada, \\
Hungarian Foundation for Scientific Research, OTKA T-038240, 
and T-042864,\\
The NWO/NATO Fund for Scientific Research, the Netherlands.\\

\bibliography{pr387}

\end{document}